\documentclass[preprint,aps]{revtex4}
\usepackage{epsf}
\usepackage{bm}% bold math
\everymath{\displaystyle}

\begin{document}

\title{Spin Oscillations in Storage Rings}

\author{A.~J. SILENKO}

\address{Institute of Nuclear Problems, Belarusian State University,\\
11 Bobruiskaya Street, Minsk 220080, Belarus\\
E-mail: silenko@inp.minsk.by}
\date{\today}

\begin{abstract}
The dependence of the particle rotation frequency on the particle
orbit perturbations is found. The exact equation of spin motion in
the cylindrical coordinate system is derived. The calculated
formula for the frequency of g-2 precession is in the best
agreement with previous results. Nevertheless, this formula
contains the additional oscillatory term that can be used for
fitting. The influence of spin oscillations on the spin dynamics
in the EDM experiment is negligible.
\end{abstract}
\maketitle
%\section{Introduction}

The goal of this investigation is obtaining ge\-ne\-ral formulae
for spin oscillations in storage rings. The problem is very
important because such oscillations change spin motion parameters.
The spin oscillations are caused by coherent betatron oscillations
(CBOs) and field distortions. In the particle rest frame, coherent
betatron oscillations lead to oscillations of magnetic field
acting on the particle and its spin. As a result, the spin
oscillates. These oscillations result in a change of the spin
rotation frequency in the g-2 experiment. If the particle
possesses an electric dipole moment (EDM), they change a vertical
spin motion measured in the EDM experiment. Perhaps, spin
oscillations can also affect other experiments.

It is quite natural to describe the spin motion in the cylindrical
coordinate system. However, the axes of this system are defined by
the position of the particle which rotates and oscillates. The
transformation of the Bargmann-Michel-Telegdi (BMT) \cite{BMT}
equation to the cylindrical coordinates should be performed with
an allowance for oscillatory terms in the particle motion
equation.

%\section{Geometrical Aspects of Particle and Spin Oscillations}

It is convenient to use the unit vector, $\bm n=\bm p/p$, which
defines the direction of particle motion.  The particle motion
equation takes the form
$$\frac{d\bm n}{dt}=\bm\omega\times\bm n,
~~~ \bm\omega=-\frac{e}{\gamma m}\left(\bm B-\frac{\bm n\times\bm
E}{\beta}\right),
$$ where $\bm\omega$
is the angular velocity of the particle rotation.

Vertical and radial CBOs change the plans of particle and spin
motion. The (pseudo)vectors of angular velocities become tilted.
The angle $\Phi$ between two positions of any rotating vector $\bm
n$ does not equal to the angle $\phi$ between two corresponding
horizontal projections. Therefore, any additional field changes
instantaneous frequencies of particle and spin motion. The average
frequencies of particle / spin motion in the tilted and horizontal
planes are the same.

The instantaneous angular velocity of particle or spin rotation in
the horizontal plane is equal to
\begin{equation}
\dot{\phi}=\frac{(\bm n_\|\times\dot{\bm n}_\|)\cdot\bm e_z}{|\bm
n_\| |^2}=\omega_3-\frac{(\omega_1n_1+\omega_2n_2)n_3}{1-n_3^2},
\label{eq1}\end{equation} where $ 1\Rightarrow\bm e_x~ {\rm or}~
\bm e_\rho, ~2\Rightarrow\bm e_y~ {\rm or}~ \bm e_\phi,~
3\Rightarrow\bm e_z$. This equation is exact.

%\section{Corrections to the Particle Motion in the Horizontal Plane}

Usually, we can take into account only perturbations of particle
orbit caused by the vertical and radial CBOs. In this case,
$$
n_1=n_\rho=p_\rho/p=\rho_0\sin{(\omega_yt+\alpha)}, ~~~ %\\
n_3=n_z=p_z/p=\psi_0\sin{(\omega_pt+\beta)}.$$

The second term in Eq. (1) is of the third order in the angular
amplitudes of oscillations, $\rho_0$ and $\psi_0$. Moreover, it
oscillates and therefore it equals zero on the average. If we take
into account only second-order terms in the angular amplitudes,
the last term in Eq. (1) is negligible. Approximately,
$\dot{\phi}=\omega_3$.
%\section{Equation of Spin Motion in Storage Rings}
The horizontal axes ($\bm e_\rho$ and $\bm e_\phi\,$) rotate with
the instantaneous angular velocity $\bm\omega'=\dot{\phi}\bm e_z.
$

Formulae for the electric dipole moment (EDM) can be obtained from
the corresponding formulae for the anomalous magnetic moment by
the substitution $\bm B\!\rightarrow\! \bm E,~ \bm
E\!\rightarrow\!-\bm B,~ g\!-\!2\!\rightarrow\!\eta$. The account
of the particle EDM leads to the following modification of the BMT
equation:
\begin{equation}\begin{array}{c}
\frac{d\bm s}{dt}=\bm\Omega\times\bm s, ~~~ \bm\Omega=\bm\Omega_{BMT}+\bm\Omega_{EDM},\\
\bm\Omega_{BMT}=-\frac{e}{m}\left[\left(a+\frac 1\gamma\right) \bm
B-\frac{a\gamma}{\gamma+1}\bm\beta(\bm\beta\cdot\bm
B)-\left(a+\frac{1}{\gamma+1}\right)\left(\bm\beta\times\bm
E\right)\right],\\ \bm\Omega_{EDM}=-\frac{e\eta}{2m}\left(\bm
E-\frac{\gamma}{\gamma+1}\bm\beta(\bm\beta\cdot\bm
E)+\bm\beta\times\bm B\right), ~~~
a=\frac{g-2}{2},\end{array}\label{eq2}\end{equation} where
$\bm\Omega$ means the angular velocity of spin rotation in the
Cartesian coordinates, and $\bm\Omega_{BMT}$ is defined by the BMT
equation. The EDM does not affect the particle motion. As a rule,
we can neglect term
$\frac{\gamma}{\gamma+1}\bm\beta(\bm\beta\cdot\bm E)$.

The spin motion equation in the cylindrical coordinate system
takes the form
\begin{equation}\begin{array}{c} \frac{ds_\rho}{dt}=\frac{d\bm s}{dt}\cdot\bm
e_\rho+\bm s\cdot\frac{d\bm e_\rho}{dt}=\bm s\cdot(\bm e_\rho\times\bm
\Omega)+\dot{\phi} s_\phi,\\
\frac{ds_\phi}{dt}=\frac{d\bm s}{dt}\cdot\bm e_\phi+\bm
s\cdot\frac{d\bm e_\phi}{dt}=\bm s\cdot(\bm e_\phi\times\bm
\Omega)-\dot{\phi} s_\rho,\\
\frac{ds_z}{dt}=\bm s\cdot(\bm e_z\times\bm\Omega).
\end{array}\label{eq3}\end{equation}

We can introduce the coordinate system with the axes $\bm e_1,\bm
e_2,\bm e_3$ corresponding to the axes $\bm e_\rho,\bm e_\phi,\bm
e_z$, respectively. For such a system, Eq. (3) can be rewritten in
the form:
\begin{equation}
 \frac{d\bm s}{dt}=\bm\omega_a\times\bm s,~~~\bm\omega_a=\bm\Omega-\dot{\phi}\bm e_3
\label{eq4}\end{equation} or
\begin{equation}\begin{array}{c}
\bm\omega_a=-\frac{e}{m}\left\{a\bm B-
\frac{a\gamma}{\gamma+1}\bm\beta(\bm\beta\cdot\bm B)
+\left(\frac{1}{\gamma^2-1}-a\right)\left(\bm\beta\times\bm
E\right)+\frac{1}{\gamma}\bigl[\bm B_\|\right.\\
\left.-\frac{1}{\beta^2}\left(\bm\beta\times\bm E\right)_\|\bigr]+
\frac{\eta}{2}\left(\bm
E-\frac{\gamma}{\gamma+1}\bm\beta(\bm\beta\cdot\bm
E)+\bm\beta\!\times\!\bm B\right)\!\right\}%\\
\!+\!\frac{(\omega_1n_1+\omega_2n_2)n_3}{1-n_3^2}\bm e_3,
\end{array}\label{eq5}\end{equation} where the sign $\|$ means the components
parallel to the $x_1x_2$-plane. Eqs. (4),(5) are exact, and
$\omega_a$ is the angular frequency of g-2 precession. These
equations describe the spin motion in an arbitrary storage ring
with an allowance for the particle and spin oscillations and the
EDM. As a rule, the last term in Eq. (5) is negligible.

%\section{Spin Oscillations in the g-2 Experiment}

In the g-2 experiment, the influence of the vertical CBO (pitch)
on the spin rotation frequency has been calculated by Farley
\cite{FPL,FQE}. The result has been confirmed by Field and
Fiorentini \cite{FF} and computer simulations.

The horizontal CBO (yaw) does not give any significant
corrections.

With the above formulae, the theory of spin oscillations in the
g-2 experiment can be developed in the very general form. The spin
motion perturbed by the vertical CBO is described by the equation
\begin{equation}\begin{array}{c} \frac{d\bm
s}{dt}=\left\{a_0+a_3\cos{[2(\omega_pt+\phi_p)]}\right\}(\bm
e_3\times\bm s)\\+
 a_2\cos{(\omega_pt+\phi_p)}(\bm e_2\times\bm s)+a_1\sin{(\omega_pt+\phi_p)}
 (\bm e_1\times\bm
s), \end{array}\label{eq6}\end{equation} where $a_1$ and $a_2$ are
first-order quantities and $a_3$ is a second-order quantity in the
angular pitch amplitude. The quantity $\omega_p$ is the angular
pitch frequency.

This equation has the exact solution.
%If only second-order terms
%in the pitch amplitude are taken into account, averaging results
%in the expression
%$$ \omega_a=a_0+
%\frac{a_0(a_1^2+a_2^2)-2a_1a_2\omega_p}{4(a_0^2-\omega_p^2)}
%+\frac{a_0(a_1^2-a_2^2)}{4(a_0^2-\omega_p^2)}\langle\cos{2(a_0
%t+\phi_0)}\rangle\cdot\frac{1+{s_3^{(0)}}^2}{1-{s_3^{(0)}}^2}.
%$$
%
Averaged g-2 frequency equals
\begin{equation}
\begin{array}{c}
\omega_a=\omega_0(1-C),  ~~~~~~~
C=\frac14\psi_0^2\left[1-\frac{\omega_0^2}{\gamma^2(\omega_0^2-\omega_p^2)}-
\frac{\omega_p^2(f-1)(f-1+2/\gamma)}{\omega_0^2-\omega_p^2}\right.\\
\left. -\frac{(\gamma-1)^2
\omega_0^2-f^2\gamma^2\omega_p^2}{\gamma^2(\omega_0^2-\omega_p^2)}
<\cos{[2(\omega_0t+\phi_0)]}>
\cdot\frac{1+{s_3^{(0)}}^2}{1-{s_3^{(0)}}^2}\right],
\end{array}\label{eq7}\end{equation}
where
$$
f=1+a\gamma-\frac{1+a}{\gamma}=1+a\beta^2\gamma-\frac{1}{\gamma}
~~~~~~~ {\rm and} ~~~~~~~ f=1+a\gamma $$ for electric and magnetic
focusing, respectively.

The average value of the last term oscillating with the frequency
$2\omega_0$ is zero. Therefore, formula (7) is in the best
agreement with previous results \cite{FPL,FQE,FF} found for the
particular case $s_3^{(0)}=0$. However, it is possible to include
the oscillatory term in a fitting process instead of its
eli\-mi\-na\-ti\-on. In the current g-2 experiment,
$f=1,~s_3^{(0)}=0,~\gamma\gg 1$, ~$\phi_0$ equals 0 or $\pi$, and
formula (7) takes the form
$$
\omega_a=\omega_0(1-C), ~~~
C=\frac14\psi_0^2\left[1-<\cos{(2\omega_0t)}>\right].$$

This formula shows the inclusion of the oscillatory term in a
fitting process is possible.

%\section{Spin Oscillations in the EDM Experiment}

In the EDM experiment, the spin motion in the horizontal plane is
strongly restricted with the radial electric field. In this case,
the spin rotation about the radial axis becomes very important
because this rotation imitates the EDM effect.

The vertical CBO leads to the spin motion described by the
equation
$$ \frac{d\bm
s}{dt}=\left\{a_0+a_3\cos{(\omega_pt+\phi_p)}\right\}(\bm
e_3\times\bm s)+ a_1\sin{(\omega_pt+\phi_p)}(\bm e_1\times\bm
 s), $$
where $1\Rightarrow\bm e_\phi,~ 2\Rightarrow\bm e_z,~
3\Rightarrow\bm e_\rho$, and $a_1$ and $a_3$ are first-order
quantities in the angular amplitude of pitch. This equation also
has the exact solution. Averaged angular frequency is given by
$$\begin{array}{c} \omega_a=a_0+
\frac{a_0a_1^2}{4(a_0^2-\omega_p^2)}\left[1+<\cos{[2(a_0t+\phi_0)]}>\cdot\frac{1+{s_3^{(0)}}^2}{1-{s_3^{(0)}}^2}\right]\\+
\frac{a_1a_3}{2\omega_p}\cdot\frac{s_3^{(0)}}{s_\|^{(0)}}<\sin{(a_0t+\phi_0)}>.
\end{array}$$

In the EDM experiment
$$\begin{array}{c}
\omega_a=\omega_0\left(1-
\frac{a^2\omega_c^2}{2\omega_p^2}\psi_0^2\right),
\end{array}$$
where $\omega_c$ is the cyclotron frequency.

The yaw correction for the EDM experiment is non\-ze\-ro:
$$\begin{array}{c} \omega_a=
\omega_0\left[1+a(\gamma-1)\gamma\left(\frac{\omega_p^2}
{2\omega_y^2}-\frac{1}{g}\right)\rho_0^2\right].
\end{array}$$

However, vertical and radial CBOs give only small corrections to
the systematical errors caused by the vertical electric field.
Since these systematical errors need to be eliminated, the
corrections for the vertical and radial CBOs are negligible.

%\section{Summary}


\begin{thebibliography}{0}

\bibitem{BMT}
V. Bargmann, L. Michel, and V. L. Telegdi, {\it Phys. Rev. Lett.}
{\bf 2}, 435 (1959).

\bibitem{FPL}
F. J. M. Farley, {\it  Phys. Lett. B} {\bf 42}, 66 (1972).

\bibitem{FQE}
F.J.M. Farley and E. Picasso, in {\em Quantum Electrodynamics},
ed. by T. Kinoshita (World Scientific, Singapore, 1990).

\bibitem{FF} J.H. Field and G. Fiorentini, Nuovo Cim. A{\bf 21},
297 (1974).

\end{thebibliography}
\end{document}